\documentstyle[eqsecnum,aps,epsf]{revtex}
\begin{document}
\title{
\begin{flushright}
{\bf IFUM 596/FT}
\end{flushright}  
Floquet Energies and Quantum Hall Effect \\
in a Periodic Potential
}
\author{ Ruggero Ferrari\thanks{Electronic address: 
ruggero.ferrari@mi.infn.it}}
\address{Dipartimento di Fisica, Universit\`a di Milano\\
via Celoria 16, 20133 Milano, Italy \\
and \\ INFN, Sezione di Milano}
\maketitle
\vspace{4mm}
\begin{abstract}  
The Quantum Hall Effect  for free electrons in external periodic
potential is discussed without using the linear response approximation.
We find that the Hall conductivity is related in a simple way
to Floquet energies (associated to the Schr\"odinger equation
in the co-moving frame).
By this relation one can analyze
the dependence of the Hall conductivity from the electric field.
Sub-bands can be introduced by the time average of the
expectation value of the Hamiltonian on the Floquet states. 
Moreover we prove previous results in form of sum rules as,
for instance: the topological character of the Hall conductivity
(being an integer multiple of $e^2/h$), the Diofantine equation
which constrains the Hall conductivity by the rational number
which measures the flux of the magnetic field through the periodicity
cell. The Schr\"odinger equation fixes in a natural way the
phase of the wave function over the reduced Brillouin zone: thus
the topological invariant providing the Hall conductivity
can be evaluated numerically without ambiguity.
\end{abstract}  
\section{Introduction and conclusions}
A milestone in the theory of the Quantum Hall Effect is the
result obtained by Thouless {\sl et al.} \cite{thou1} concerning
the Hall conductivity $\sigma_{\rm H}$ for free electrons 
in a periodic potential. Their proof showed that 
$\sigma_{\rm H}$ is a multiple integer
of $e^2/h$, if the chemical potential lies in a gap of the
Hofstadter spectrum \cite{hof} (gap condition). 
The result is rather striking, since
the coefficient varies strongly with the number of filled
bands and with the commensurability factor
$q/p$, which gives the flux of the magnetic field through the
periodicity cell $\Phi= {p\over q}{{hc}\over{e}}$. Subsequent
works 
\cite{avron1,simon,kohmoto1,niu1,niu2,avron2,bellissard1,imai}
have shown the generality of the result by evidencing
the topological nature of the Hall conductivity. 
Moreover the result has been shown to be valid also in presence
of many-body interaction, provided the ground state is not
degenerate  \cite{niu1,niu2,avron2}
\par
The proof by Thouless {\sl et al.} makes use 
of the linear response approximation
(Kubo formula). To our knowledge all works devoted to
this problem use the linear approximation or, equivalently,
the adiabatic approximation. 
\par
There are few reasons that make the case of {\sl finite}
electric field an interesting problem. First, from the experimental
point of view it may be interesting to observe the phenomenon
when the electric field varies \cite{gerhardts}. 
Second, the limit of weak periodic
potential is in conflict with the limit of small electric field.
It is important to investigate the intermediate situation where
the potential is weak in comparison to the Landau splitting
and comparable to the electrostatic potential.
\par
The present work deals with the problem without using the
linear response approximation. We consider a Galilei transformation
in order to have the time dependence of the Hamiltonian
only in the external periodic potential. 
The Hamiltonian, being periodic both under magnetic and time
translations, allows an analysis in terms of
Bloch functions and Floquet eigenstates \cite{bellissard2}. 
Thus the dynamical problem consists in finding
the quasi-periodic solutions of the Schr\"odinger equation
(Floquet states) and their quasi-energies (Floquet energies).
\par
It is shown that there
is a simple relation between the Hall conductivity and
the Floquet energies  
associated to the periodic Hamiltonian.
Floquet energies can be easily obtained by diagonalization the
time evolution operator. The average in time over a period of
the expectation value of the Hamiltonian over Floquet states
turns out to be a convenient quantity in order to define 
{\sl sub-bands} on the reduced Brillouin zone.
A strong electric field complicates the construction of
a stationary state, since it induces transitions to energies
above the chemical potential. In fact a spectrum of energy,
which satisfies the gap condition in the adiabatic approximation,
might appear partially unresolved for finite values of the electric field.
\par
If the gap condition is satisfied, then the Hall conductivity
is an integer and moreover it is ruled by a Diofantine
equation \cite{thou1,dana,kohmoto2,kohmoto3}. 
These results are valid as sum rules, i.e. in the
case where a group of sub-bands cross with each other.
\par
In the adiabatic approximation there is a simple relation
between Floquet energies and Berry phases. Moreover the
phase of the instantaneous eigenvector can be chosen
in an essentially unique way \cite{kohmoto1}. Thus the Hall conductivity
can be evaluated by means of a suitable line integral
on the border of the reduced Brillouin zone $\cal C$ \cite{thou1,berry}
\begin{eqnarray}
-  {i\over{2 }}\int_{\partial{\cal C}}dw_l
\big(
\langle{\partial\over{\partial w_l}} \Upsilon^{(w)}_J|
\Upsilon^{(w)}_J\rangle -
\langle \Upsilon^{(w)}_J|{\partial\over{\partial w_l}}
\Upsilon^{(w)}_J\rangle 
\big)
\label{paper4.57p}
\end{eqnarray}
(see eq. \ref{paper4.57}).
\section{The model}
The Schr\"odinger equation in presence of
a magnetic field $B$ (along $z$, in the symmetric gauge)
and of an electric field $\bf E$ is \cite{notations}
\begin{eqnarray}
&&i\partial_t\psi = H\psi
\nonumber \\
&&
H={1\over 2} \big [   (-i\partial_1 - 
{{r_2}\over 2} - {\cal E}_1t)^2 +
(-i\partial_2 + {{r_1}\over 2} - 
{\cal E}_2t)^2  \big ]+ {\cal V}({\bf r})
\label{paper4.1}
\end{eqnarray}
where 
\begin{eqnarray}
{\cal E}_i = {{e\lambda E_i}\over{\hbar\omega}},
\qquad
{\cal V}({\bf r}) = {1\over{\hbar\omega}} V(\lambda {\bf r}).
\label{paper4.2}
\end{eqnarray}
The external potential is periodic over a lattice 
\begin{eqnarray}
{\cal V}({\bf r}+m{\bf c'}+n{\bf d'}) =  {\cal V}({\bf r})
\quad
\forall (m,n)\in {\cal Z}^2.
\label{paper4.3}
\end{eqnarray}
The flux through the cell 
is given by a rational number of the quantum unit of flux
\begin{eqnarray}
{\tilde{\bf c}}'\cdot{\bf d}' = 2\pi {p\over{q}}.
\label{paper4.4}
\end{eqnarray}
Thus we introduce a convenient sub-lattice $({\bf c},{\bf d})$
and a finite domain with area $A$
and side-vectors $({\bf L}_1,{\bf L}_2)$ 
where the electrons live.
The commensurability is given by
\begin{equation}
\left \{
\begin{array}{l}
{\bf c}=r{\bf c}'+s'{\bf d}'
\\
{\bf d}=r' {\bf c}' + s{\bf d}'
\\
r,s,r',s'~~{\rm integers}
\\
rs-r's'=q
\end{array}
\right .
\quad
\left \{
\begin{array}{l}
{\bf L}_1 = k {\bf c}+ l' {\bf d}
\\
{\bf L}_2 = k' {\bf c}+ l {\bf d}
\\
k,l,k',l'~~{\rm integers}
\\
kl-k'l' = N \equiv {{g_L}\over p}
\end{array}
\right .
\label{paper4.5}
\end{equation}
where $g_L= A/(2\pi)$ is the degeneracy of the (unperturbed) Landau
level.
The boundary conditions for eq. (\ref{paper4.1}) are
imposed by means of the Magnetic Translation operator 
\cite{ferrari,imai}
\begin{equation}
S(v) \equiv \exp({i\over {2}}{\tilde{\bf v}}
\cdot{\bf r})\exp(v_i\partial_i)
\label{paper4.6}
\end{equation}
over the domain of definition given by the parallelogram
with side vectors ${\bf L}_1,{\bf L}_2$.
We consider periodic boundary conditions
\begin{eqnarray}
S(L_1)\psi &=& e^{i\theta_1}\psi
\nonumber \\
S(L_2)\psi &=& e^{i\theta_2}\psi.
\label{paper4.7}
\end{eqnarray}
\par
Both operators $S(c)$ and $S(d)$ commute with
the Hamiltonian and moreover \cite{allowed}
\begin{eqnarray}
[S(c),S(d)]=0.
\label{paper4.9}
\end{eqnarray}
Then the solutions of eq. (\ref{paper4.1}) can be labeled
by the phases $\mu,\nu$ given by 
\begin{eqnarray}
S(c)\psi^{\mu\nu} &=& e^{i\mu}\psi^{\mu\nu}
\nonumber \\
S(d)\psi^{\mu\nu} &=& e^{i\nu}\psi^{\mu\nu}.
\label{paper4.10}
\end{eqnarray}
The values of $\mu,\nu$ are fixed by the conditions
(\ref{paper4.5}) and (\ref{paper4.7}) in a standard way.
\section{Galilei transformation and Bloch functions}
We consider a Galilei transformation which removes the
electric field from the kinetic term in eq. (\ref{paper4.1}).
We use the operator \cite{ferrari}
\begin{equation}
T(v) \equiv \exp(-{i\over {2}}{\tilde{\bf v}}
\cdot{\bf r})\exp(v_i\partial_i)
\label{paper4.11}
\end{equation}
which commutes with any operator $S$. The unitary
transformation
\begin{equation}
\psi^{(G)} =  T(-({\cal E}+ {\tilde{\cal E}}t))\psi     
\label{paper4.12}
\end{equation}
yields a function which satisfies a Schr\"odinger equation
with Hamiltonian
\begin{eqnarray}
H^{(G)} &=& 
{1\over 2} \big [   (-i\partial_1 - 
{{r_2}\over 2} )^2 +
(-i\partial_2 + {{r_1}\over 2} )^2
\big ]\nonumber \\
&&+ {\cal V}( r-({\cal E}+ {\tilde{\cal E}}t)).
\label{paper4.13}
\end{eqnarray}
$\psi^{(G)}$ can be chosen to satisfy the conditions
(\ref{paper4.10}).
\par
The phases $\mu,\nu$ assume a finite number of values, proportional
to the area $A$ of the domain. Eventually we will extrapolate to
continuous values. However this cannot be done directly on $\psi^{\mu\nu}$
(eq. (\ref{paper4.10})), without running into the following paradox.
Take the derivative respect
to $\mu$ and then the expectation value on $\psi^{\mu\nu}$
\begin{eqnarray}
\langle \psi^{\mu\nu}, S(c){\partial\over{\partial\mu}}\psi^{\mu\nu}\rangle
= ie^{i\mu}\Vert \psi^{\mu\nu}\Vert^2 + e^{i\mu}
\langle \psi^{\mu\nu},{\partial\over{\partial\mu}}\psi^{\mu\nu}\rangle.
\label{paper4.20a}
\end{eqnarray}
Then use the  unitarity for $S(c)$ 
\begin{eqnarray}
\langle S(-c) \psi^{\mu\nu},{\partial\over{\partial\mu}}
\psi^{\mu\nu}\rangle
= e^{i\mu}
\langle \psi^{\mu\nu},{\partial\over{\partial\mu}}
\psi^{\mu\nu}\rangle.
\label{paper4.20b}
\end{eqnarray}
From eqs. (\ref{paper4.20a}) and (\ref{paper4.20b}) one gets
\begin{equation}
\Vert \psi^{\mu\nu}\Vert = 0.
\label{paper4.20c}
\end{equation}
This result is absurd.
\par
This difficulty can be avoided by introducing the unitary
equivalent functions
\begin{equation}
\Upsilon^{(w_{\mu\nu})} \equiv
S(-w_{\mu\nu})\psi^{(G)\mu\nu}
\label{paper4.14}
\end{equation}
where
\begin{equation}
w_{\mu\nu} = {1\over{2\pi p}}
[(\nu-\nu_0)c   -(\mu-\mu_0) d]
\label{paper4.15}
\end{equation}
($\mu_0,\nu_0$ are fixed).
The corresponding Hamiltonian is
\begin{equation}
H^{(\Upsilon)}_{w_{\mu\nu}} ={1\over 2} 
\big [   (-i\partial_1 - 
{{r_2}\over 2} )^2 +
(-i\partial_1 + {{r_1}\over 2} )^2  \bigr]
+ {\cal V}(r-w_{\mu\nu} -({\cal E} + {\tilde{\cal E}}t)).
\label{paper4.17}
\end{equation}
$\Upsilon^{(w_{\mu\nu})}$ are like Bloch functions. In fact,
by using the composition law
of $S$
\begin{equation}
S(v) S(w)= S(v+w) \exp(-{i\over {2}}{\tilde{\bf v}}
\cdot{\bf w}),
\label{paper4.19}
\end{equation}
one can easily check that
\begin{eqnarray}
S(c)\Upsilon^{(w_{\mu\nu})} &=& e^{i\mu_0}\Upsilon^{(w_{\mu\nu})}
\nonumber \\
S(d)\Upsilon^{(w_{\mu\nu})} &=& e^{i\nu_0}\Upsilon^{(w_{\mu\nu})},
\label{paper4.20}
\end{eqnarray}
i.e. the boundary conditions on the periodicity lattice
are fixed. This fact makes the extrapolation to 
continuous values of $w_{\mu\nu}$ harmless. In particular the unitarity
property of operators $S$ is preserved.
\par
The reduced Brillouin zone is given by the domain containing 
all possible values of $w_{\mu\nu}$
\begin{equation}
{\cal C}\equiv\left\{ {\bf w}\in {\cal R}^2:
{\bf w} = {1\over p}(\lambda_1 {\bf c}+\lambda_2 {\bf d}), \quad
0<\lambda_j<1 \quad j=1,2 \right \}.
\label{paper4.21}
\end{equation}
\par
If $\tau {\tilde{\cal E}}$ is a site of the 
lattice ${\bf c'},{\bf d'}$,
the Hamiltonian $H^{(\Upsilon)}_{w_{\mu\nu}}$ is periodic
in time with period $\tau $. Thus we assume
\begin{equation}
\tau {\tilde{\cal E}}   = m_0 {\bf c}' + n_0 {\bf d}' \qquad
(m_0,n_0~{\rm integers}).
\label{paper4.22}
\end{equation}
\par
For technical reasons we introduce another set of 
(unitary equivalent) functions and the corresponding Hamiltonian
($T(w)$ exists for any vector $w$ \cite{allowed})
\begin{equation}
\Xi^{(w)} \equiv T(w) \Upsilon^{(w)}
\label{paper4.16}
\end{equation}
\begin{equation}
H^{(\Xi)}_{w} ={1\over 2} 
\big [   (-i\partial_1 - 
{{r_2}\over 2}-w_2 )^2 +
(-i\partial_1 + {{r_1}\over 2}+w_1 )^2  \bigr]
+ {\cal V}(r -({\cal E} + {\tilde{\cal E}}t)).
\label{paper4.18}
\end{equation}
%
\section{The Hall current}
The stationary state for a system with time-dependent
Hamiltonian at zero temperature (in the present
case it is a good approximation) can
be defined by a minimum criterion. For any
change on the stationary state 
\begin{equation}
\Delta\left\{{1\over {2t_0}}
\int_{-t_0}^{t_0} dt \langle H^{(\Upsilon)}(t)\rangle 
- \epsilon_F \langle N_e\rangle \right\}\geq 0
\label{paper4.23}
\end{equation}
with $t_0>>\tau $ and
where $\epsilon_F$ is the chemical potential and $N_e$ is
the operator that counts the number of electrons.
\par
The space average
of the Hall current is (in units $e\hbar/(m\lambda^3)$)
\begin{eqnarray}
J_1 &=&{1\over{(2\pi)^2 }}
\int_{\cal C}d^2 w \sum_J' 
\langle \Upsilon^{(w)}_J|[p_1 - 
{{r_2}\over 2} + {\cal E}_2
]| \Upsilon^{(w)}_J\rangle
\nonumber \\
J_2 &=&{1\over{(2\pi)^2 }}
\int_{\cal C}d^2 w \sum_J' 
\langle \Upsilon^{(w)}_J|[p_2 + 
{{r_1}\over 2} - {\cal E}_1
]| \Upsilon^{(w)}_J\rangle.
\label{paper4.24}
\end{eqnarray}
The index $J$ labels a complete set of normalized solutions
of the Schr\"odinger equation. In Sections \ref{sec:floquet} 
and \ref{sec:sum} we
provide a choice for the orthonormal set by constructing sub-bands.
The conditioned sum is over those states which satisfy the
condition in eq. (\ref{paper4.23}). In general the sum
over $J$ is discontinuous in $w$.  In terms of $\Xi$ functions the current is 
\begin{equation}
J_i = -\epsilon_{ij}
{1\over{(2\pi)^2 }}\int_{\cal C}d^2 w\sum_J'
\langle \Xi^{(w)}_J|
{{\partial H^{(\Xi)}_w}\over{\partial w_j}}
| \Xi^{(w)}_J\rangle
+ {{\nu_f}\over{2\pi}}\epsilon_{ij}{\cal E}_j
\label{paper4.25}
\end{equation}
where $\nu_f$ is the filling factor.
By using the Schr\"odinger equation it can be written as
\begin{eqnarray}
J_i= 
-i\epsilon_{ij}
{1\over{(2\pi)^2 }}\int_{\cal C}d^2 w\sum_J'
\partial_t[
\langle \Xi^{(w)}_J|\partial_j
| \Xi^{(w)}_J\rangle] +
{{\nu_f}\over{2\pi}}\epsilon_{ij}{\cal E}_j.
\label{paper4.26}
\end{eqnarray}
The dependence from wave function $\Upsilon^{(w)}_J$ 
can be obtained
by introducing the creation and annihilation operators
$a,a^\dagger$
\begin{eqnarray}
&&\partial_1T(w) =T(w)
\Big[
{1\over\sqrt{2}}(a-a^\dagger)+{i\over 2}w_2
\Big]
\nonumber\\
&&\partial_2T(w) =T(w)
\Big[
{i\over\sqrt{2}}(a+a^\dagger)- {i\over 2}w_1
\Big].
\label{paper4.27}
\end{eqnarray}
We get
\begin{eqnarray}
J_1 &=& {{\nu_f}\over{2\pi}}{\cal E}_2
+
{1\over{(2\pi)^2 }}\int_{\cal C}d^2 w\sum_J'
\partial_t\Big\{
\langle \Upsilon^{(w)}_J|
[{1\over\sqrt{2}}(a+a^\dagger)
-{1\over 2}w_1]|\Upsilon^{(w)}_J\rangle
\nonumber \\
&-&i\langle \Upsilon^{(w)}_J|\partial_2
\Upsilon^{(w)}_J\rangle \Big\}
\nonumber \\
J_2 &=& -{{\nu_f}\over{2\pi}}{\cal E}_1
+
{1\over{(2\pi)^2 }}\int_{\cal C}d^2 w\sum_J'
\partial_t\Big\{
\langle \Upsilon^{(w)}_J|
[{i\over\sqrt{2}}(a-a^\dagger)
-{1\over 2}w_2]|\Upsilon^{(w)}_J\rangle
\nonumber \\
&+&
i\langle \Upsilon^{(w)}_J|\partial_1
\Upsilon^{(w)}_J\rangle \Big\}.
\label{paper4.28}
\end{eqnarray}
We consider now the time average of the density of current over a period.
If the wave functions obey the Floquet condition
\begin{equation}
\Upsilon^{(w)}_J(t+\tau ) = \exp\{-iE^{(w)}_J\}
\Upsilon^{(w)}_J(t)
\label{paper4.29}
\end{equation}
then
\begin{equation}
\langle J_i\rangle =\epsilon_{ij}{1\over{2\pi}}\Big [
\nu_f{\cal E}_j
-i 
{1\over{2\pi }}\int_{\cal C}d^2 w \sum_J'
{1\over \tau }\int_0^{\tau }dt\partial_t
\langle \Upsilon^{(w)}_J|\partial_j
\Upsilon^{(w)}_J\rangle 
\Big ].
\label{paper4.30}
\end{equation}
Now we take the components parallel and orthogonal 
to the electric field
and get the Hall conductivity (in units $e^2/h$)
\begin{eqnarray}
\sigma_{xx} &\equiv& {{\cal E}_i\over {{\cal E}^2}}
\langle J_i\rangle
\nonumber \\
& =&-i\epsilon_{ik}{{\cal E}_i\over {{\cal E}^2}} 
{1\over{2\pi }}\int_{\cal C}d^2 w \sum_J'
{1\over \tau }\int_0^{\tau }dt\partial_t
\langle \Upsilon^{(w)}_J|\partial_k
\Upsilon^{(w)}_J\rangle 
\label{paper4.31}
\\
\sigma_{xy} &\equiv&
\epsilon_{ik}{{\cal E}_k\over {{\cal E}^2}}
\langle J_i\rangle
\nonumber \\
& =&\Big [\nu_f
-
{i\over{2\pi }}\int_{\cal C}d^2 w \sum_J'
{1\over \tau }\int_0^{\tau }dt\partial_t
{{\cal E}_k\over {{\cal E}^2}}
\langle \Upsilon^{(w)}_J|\partial_k
\Upsilon^{(w)}_J\rangle 
\Big ].
\label{paper4.31a}
\end{eqnarray}
Since eq. (\ref{paper4.29}) is valid for any $t$, the Hall
conductivity in eqs. (\ref{paper4.31}) and (\ref{paper4.31a})  
is constant in time.
\section{Floquet energy}
\label{sec:floquet} 
The time evolution of the wave functions $\Upsilon^{(w)}$
is given by the Schr\"odinger equation associated to the Hamiltonian
in eq. (\ref{paper4.17}) with the boundary conditions 
(\ref{paper4.20}). Consider the unitary evolution operator 
$U^{(w)}(t,t')$ which satisfies
\begin{equation}
i\partial_t U^{(w)}(t,t')= H^{(\Upsilon)}_w U^{(w)}(t,t') 
\qquad U^{(w)}(t',t')=1.
\label{paper4.32}
\end{equation}
The solution of eq. (\ref{paper4.32}) is
\begin{equation}
U(t,s) = \sum_{n=0,\infty}{{(-i)^n}\over{n!}}
\int_s^tdt_1\dots dt_n T\left(
H^{(\Upsilon)}_w(t_1)\dots H^{(\Upsilon)}_w(t_n)\right).
\label{paper4.32a}
\end{equation}
The solutions of the Schr\"odinger equation, that
are quasi-periodic in time (eq. (\ref{paper4.29})), are
both right and (complex conjugate) left eigenvectors
of  $U^{(w)}(t'+\tau ,t')$. The eigenvalues
\begin{equation}
\exp(-i E^{(w)}_J )
\label{paper4.33}
\end{equation}
provide the Floquet energies $E^{(w)}_J ~[{\rm mod}(2\pi)]$ 
\cite{bellissard2}.
\par
In the adiabatic  limit the Floquet energies are given by
\begin{equation}
E_J^{(w)} = \int_0^\tau  dt \lambda_J(w+{\tilde{\cal E}}t)
+ \alpha_J^{(w)}
\label{paper4.44}
\end{equation}
where $\lambda_J(w)$ is the instantaneous eigenvalue
of the Hamiltonian and
$\alpha_J^{(w)}$ is the Berry phase \cite{thou1,berry}. 
\par
One can easily verify that
\begin{equation}
S(-{c\over p})U^{(w)}(t'+\tau ,t')
S({c\over p})=U^{(w+{c\over p})}(t'+\tau ,t'),
\label{paper4.34}
\end{equation}
then the set of eigenvalues is the same
\begin{equation}
\left\{\exp(-i E^{(w)}_J )\right\}=
\left\{
\exp(-i E^{(w+{c\over p})}_J )
\right\} \quad\forall w.
\label{paper4.35}
\end{equation}
Similarly one gets
\begin{equation}
\left\{\exp(-i E^{(w)}_J )\right\}=
\left\{
\exp(-i E^{(w+{d\over p})}_J )
\right\} \quad\forall w.
\label{paper4.36}
\end{equation}
Thus we get
\begin{eqnarray}
&&
E^{(w)}_J = E^{(w+{c\over p})}_{J_c}+ 2\pi k_{Jc}
\nonumber\\
&&
E^{(w)}_J =
E^{(w+{d\over p})}_{J_d} + 2\pi k_{Jd}
\quad\forall w
\label{paper4.37}
\end{eqnarray}
where $k_{Jc},k_{Jd}$ are integers. The one-to-one
mapping $J\to J_c$ and $J\to J_d$ depends
strongly on the direction and on the strength of
the electric field.
A further consequence
of eq. (\ref{paper4.34}) is that 
\begin{eqnarray}&&
\Upsilon^{(w)}_J = \exp(i\phi_{J}(w))
S({c\over p})\Upsilon^{(w+{c\over p})}_{J_c}
\nonumber\\
&&
\Upsilon^{(w)}_J= \exp(i\psi_{J}(w))
S({d\over p})\Upsilon^{(w+{d\over p})}_{J_d}
\quad\forall w
\label{paper4.38}
\end{eqnarray}
where $\phi_{J}(w), \psi_{J}(w)$ are phases.
\par
The periodic time dependence of the Hamiltonian
gives further informations about the Floquet
energies
\begin{eqnarray}
U^{(w)}(t-t',s-t') = U^{(w+{\tilde{\cal E}}t')}(t,s).
\label{paper4.39}
\end{eqnarray}
Therefore the Floquet energies are constant along
the line $w+{\tilde{\cal E}} t$
\begin{equation}
{\partial\over {\partial t}}
E^{(w+{\tilde{\cal E}} t)}_J = 0.
\label{paper4.40}
\end{equation}
This implies that the longitudinal Hall conductivity
$\sigma_{xx}$ in eq. (\ref{paper4.31}) is zero. Moreover
for the transverse Hall conductivity the reduced 
Brillouin zone can be chosen to be a rectangular ${\cal C}'$
with  side vectors laying on ${\cal E}$ and $\tilde{\cal E}$.
Either ${\bf c}/p$ (or ${\bf d}/p$) lies on the side 
${\cal E}$. Then a simple inspection to the geometry
gives
\begin{eqnarray}
\sigma_{xy}
& =&\Big [
\nu_f
-
{1\over{\tau{\bf c}\cdot {\cal E} }}
\int_\Gamma d w_k\sum_J'
\partial_k E^{(w)}_J
\Big ]
\label{paper4.41}
\end{eqnarray}
with 
\begin{eqnarray}
\tau{\bf c}\cdot {\cal E} = 2\pi{{p}\over{q}}
(n_0 r-m_0s').
\label{paper4.41x}
\end{eqnarray}
The line integral $\Gamma$ is the straight segment along the electric
field, starting in $p^{-1}[{\bf c} -
{\cal E}^{-2}({\bf c}\cdot {\cal E} ){\cal E}]$ and ending in
${\bf c}/p$. Equation
(\ref{paper4.41}) provides a direct way to evaluate the Hall
conductivity. One has to find the Floquet energies and then
the sub-bands have to be reconstructed by requiring continuity
on the reduced Brillouin zone. If the gap condition is satisfied
then any sub-band is either filled or empty. As a consequence
of this situation we get 
\begin{eqnarray}
\int_\Gamma d w_k\sum_J'
\partial_k E^{(w)}_J=\sum_J'\int_\Gamma d w_k\partial_k E^{(w)}_J =
\sum_J'\Big( E^{({c\over p})}_J-E^{(0)}_J\Big).
\label{paper4.41xp}
\end{eqnarray}
\par
The periodicity of the Hamiltonian in $w$ across $\bf c', d'$
in eq. (\ref{paper4.17}) implies relations similar to those
in eqs. (\ref{paper4.35}) and (\ref{paper4.36})
\begin{eqnarray}
&&
\left\{\exp(-i E^{(w)}_J )\right\}=
\left\{
\exp(-i E^{(w+{c'})}_J )
\right\} 
\nonumber\\
&&
\left\{\exp(-i E^{(w)}_J )\right\}=
\left\{
\exp(-i E^{(w+{d'})}_J )
\right\} \quad\forall w.
\label{paper4.41a}
\end{eqnarray}
They imply
\begin{eqnarray}
&&
E^{(w)}_J = E^{(w+{c'})}_{J'_c}+ 2\pi k'_{Jc}
\nonumber\\
&&
E^{(w)}_J =
E^{(w+{d'})}_{J'_d} + 2\pi k'_{Jd}
\quad\forall w.
\label{paper4.41b}
\end{eqnarray}
Again the mapping among Floquet energies is generally
non-trivial.
\section{Sub-bands, clusters and sum rules}
\label{sec:sum}
The sub-bands are surfaces given by
\begin{eqnarray}
{\cal I}_J^{(w)} \equiv {1\over\tau}\int_0^\tau dt 
\langle \Upsilon_J^{(w)}(t)|H_w^{(\Upsilon)}(t)|
\Upsilon_J^{(w)}(t)\rangle 
\label{paper4.41ab}
\end{eqnarray}
where $w$ is any point in the reduced Brillouin zone
and $J$ denotes the sub-band. The raw data given
by the eigenvalues and eigenvectors of $U(\tau,0)$
have to be organized so that the sub-bands are regular
surfaces. Figs. 1-10 give examples of this procedure.
Although for a finite strength of the 
electric field the sub-bands described
by ${\cal I}_J{(w)}$ are complicated surfaces which cross
each other,
in general the construction of a sub-band shows no difficulties
since the Floquet energies $E_J{(w)}$ appear to be
smooth functions. Some properties are helpful
for their construction.
Eq. (\ref{paper4.40}) tells that sub-bands vary only in the
direction given by the electric field. Moreover eqs.
(\ref{paper4.37}) and (\ref{paper4.38}) give bounds on the
boundary of the reduced Brillouin zone. As expected, in the
adiabatic limit one observes the eigenvalue patterns of the
instantaneous Hamiltonian (their time averages). 
As the strength of the electric field rises, the mixing
of these eigenvalues to form a sub-band becomes more and more
complicated. Sometime the construction of the sub-band
is difficult due to a pinching process on two (or more) lines.
This process is exemplified in Figs. 5 and 9. In particular
Fig. 9 shows that as the electric field decreases the v-like
lines merge to form x-crossing regular lines. At the same time
a gap appears as in Fig. 3.
\par
We consider here the situation where the sub-bands flock in
clusters separated by gaps. Then
the mapping of the Floquet energies in $w$ to those in
$w+ {c\over p}$, $w+ {d\over p}$, $w+c'$ and $w+d'$ is limited
to the set separated by gaps. A  sub-band has the time average
of the energy that crosses at least one of the other elements
of the set, but none of those outside. Thus eqs. (\ref{paper4.37})
and (\ref{paper4.41b}) provide the sum rules for every cluster
$\cal S$
\begin{eqnarray}
&&
\sum_{J\in {\cal S}}E^{(w+{c\over p})}_J =
\sum_{J_c\in {\cal S}}E^{(w)}_{J_c} + 2\pi {\cal K}_{{\cal S}c}
\qquad {\cal K}_{{\cal S}c}\equiv \sum_{J\in {\cal S}}k_{Jc}
\nonumber\\
&&
\sum_{J\in {\cal S}}E^{(w+{d\over p})}_J =
\sum_{J_d\in {\cal S}}E^{(w)}_{J_d} + 2\pi {\cal K}_{{\cal S}d}
\qquad {\cal K}_{{\cal S}d}\equiv \sum_{J\in {\cal S}}k_{Jd}
\label{paper4.37p}
\end{eqnarray}
and
\begin{eqnarray}
&&
\sum_{J\in {\cal S}}E^{(w+{c'})}_J =
\sum_{J'_c\in {\cal S}}E^{(w)}_{J'_c} + 2\pi {\cal K}'_{{\cal S}c}
\qquad {\cal K}'_{{\cal S}c}\equiv \sum_{J\in {\cal S}}k'_{Jc}
\nonumber\\
&&
\sum_{J\in {\cal S}}E^{(w+{d'})}_J =
\sum_{J'_d\in {\cal S}}E^{(w)}_{J'_d} + 2\pi {\cal K}'_{{\cal S}d}
\qquad {\cal K}'_{{\cal S}d}\equiv \sum_{J\in {\cal S}}k'_{Jd}.
\label{paper4.41bp}
\end{eqnarray}
The contribution of a filled set of sub-bands to the Hall conductivity
is then by eq.  (\ref{paper4.41xp})
\begin{eqnarray}
\sigma_{xy}
& =&
\nu_f
-{{2\pi}\over{\tau{\bf c}\cdot {\cal E} }}
\sum_{{\cal S}}'{\cal K}_{{\cal S}c}.
\label{paper4.41c}
\end{eqnarray}
The periodicity in time expressed by eq. (\ref{paper4.22})
implies (see also eq. (\ref{paper4.40}))
\begin{equation}
m_0 {\cal K}'_{{\cal S}c}+ n_0  {\cal K}'_{{\cal S}d} = 0.
\label{paper4.41bb}
\end{equation}
Since $m_0, n_0$ are relative prime numbers then an integer 
${\cal K}'_{\cal S}$ exists such that
\begin{eqnarray}
&& {\cal K}'_{{\cal S}c} = n_0 {\cal K}'_{\cal S}
\nonumber \\
&&  {\cal K}'_{{\cal S}d} = -m_0 {\cal K}'_{\cal S}.
\label{paper4.41cc}
\end{eqnarray}
Moreover the commensurability relations in eqs. (\ref{paper4.5})
give
\begin{eqnarray}
&& {\cal K}_{{\cal S}c} = 
{1\over p}[r {\cal K}'_{{\cal S}c} + s'{\cal K}'_{{\cal S}d}]=
{{{\cal K}'_{\cal S}}\over p}[r n_0 - s'  m_0]
\nonumber \\
&&  {\cal K}_{{\cal S}d} = 
{1\over p}[r' {\cal K}'_{{\cal S}c} + s  {\cal K}'_{{\cal S}d}]=
{{{\cal K}'_{\cal S}}\over p}[r' n_0 - s  m_0].
\label{paper4.41d}
\end{eqnarray}
\par
Finally we get by using eq. (\ref{paper4.41x})
\begin{eqnarray}
\sigma_{xy}
={1\over{p}}\Big [
n_f
-\sum_{\cal S}'
{{q}\over{p}}{\cal K}'_{{\cal S}}
\Big ]
\label{paper4.41e}
\end{eqnarray}
where $n_f$ is the number of filled sub-bands.
\section{The topological invariant}
The expression of $\sigma_{\rm H}$ as a topological
invariant derived by Thouless {\sl et al.} can be obtained
from eq. (\ref{paper4.31a}) by using the sole hypothesis
that the chemical potential lies in a gap of the sub-bands.
The conditioned sum implies that in such a case 
any sub-band is either filled
or empty. In this situation, by using the periodicity
relations in eq. (\ref{paper4.38}), one gets
\begin{equation}
\int_{\cal C}d^2 w
\partial_k\langle \Upsilon^{(w)}_J|\partial_t
\Upsilon^{(w)}_J\rangle = 0.
\label{paper4.48}
\end{equation}
Then the Hall resistance (\ref{paper4.31a}) can be written
\begin{eqnarray}
&&\sigma_{xy} 
=\Big [
{{n_f}\over{p}} -
\nonumber\\
&&
{i\over{2\pi }}\sum_J'\int_{\cal C}d^2 w
{1\over \tau }\int_0^{\tau }dt
{{\cal E}_k\over {{\cal E}^2}}\big(
\partial_t
\langle \Upsilon^{(w)}_J|\partial_k
\Upsilon^{(w)}_J\rangle -
\partial_k
\langle \Upsilon^{(w)}_J|\partial_t
\Upsilon^{(w)}_J\rangle 
\big)
\Big ].
\label{paper4.49}
\end{eqnarray}
Since $\Upsilon$ depends from $w$ and $t$ only through
$w+{\tilde{\cal E}}t$, then $\partial_t={\tilde{\cal E}}_j\partial_j$.
Finally one gets the Hall conductivity as a topological invariant
\cite{thou1}
\begin{eqnarray}
&&\sigma_{xy} 
=\Big [
{{n_f}\over{p}} -
\nonumber\\
&& 
{i\over{2\pi }}\sum_J'\int_{\cal C}d^2 w
\big(
\langle\partial_2 \Upsilon^{(w)}_J|\partial_1
\Upsilon^{(w)}_J\rangle -
\langle\partial_1 \Upsilon^{(w)}_J|\partial_2
\Upsilon^{(w)}_J\rangle 
\big)
\Big ].
\label{paper4.50}
\end{eqnarray}
The expression in Ref. \cite{thou1} does not contain
the first term of the RHS since it is written in the reference
frame where the periodic potential is at rest.
We prefer the use of eq. (\ref{paper4.50}) since
the corresponding Hamiltonian is explicitly periodic
in $w$. The proof that the RHS of eq. (\ref{paper4.50}) is
an integer is similar to the argument given in Ref. \cite{thou1},
with the clause of considering all the sub-bands belonging to a
cluster.
We use the relations in eqs. (\ref{paper4.38}). 
By imposing monodromy after a tour along the border of the
reduced Brillouin zone, one gets
\begin{eqnarray}
-\psi_J(0)-\phi_{J_d}({d\over p})
+\psi_{J_c}({c\over p})+\phi_J(0) = 
{{2\pi}\over p} -  2\pi k_{J\sigma}
\label{paper4.55}
\end{eqnarray}
where $k_{J\sigma}$ is an integer.
Then the conductivity
$\sigma_{xy}$ is (via Stoke's theorem)
\begin{eqnarray}
&&
\sigma_{xy}= 
{{n_f}\over{p}} - 
{i\over{4\pi }}\sum_J'\int_{\partial{\cal C}}dw_l
\big(
\langle\partial_l \Upsilon^{(w)}_J|
\Upsilon^{(w)}_J\rangle -
\langle \Upsilon^{(w)}_J|\partial_l
\Upsilon^{(w)}_J\rangle 
\big)
\nonumber\\
&&= 
{{n_f}\over{p}} - {i\over{4\pi }}\sum_{\cal S}'\sum_{J\in{\cal S} }
\Big \{ \int_{w\in[0,{c\over p}]}dw_l 
\big [i\partial_l\psi_J(w) 
+ \big(
\langle\partial_l \Upsilon^{(w+{d\over p})}_{J_d}|
\Upsilon^{(w+{d\over p})}_{J_d}\rangle
\nonumber\\
&&
-\langle \Upsilon^{(w+{d\over p})}_{J_d}|\partial_l
\Upsilon^{(w+{d\over p})}_{J_d}\rangle 
\big)
-\big(
\langle\partial_l \Upsilon^{(w+{d\over p})}_J|
\Upsilon^{(w+{d\over p})}_J\rangle
-\langle \Upsilon^{(w+{d\over p})}_J|\partial_l
\Upsilon^{(w+{d\over p})}_J\rangle 
\big)
\big]
\nonumber\\
&&
+\int_{w\in[0,{d\over p}]}dw_l
\big [-i\partial_l\phi_J(w)
- \big(
\langle\partial_l \Upsilon^{(w+{c\over p})}_{J_c}|
\Upsilon^{(w+{c\over p})}_{J_c}\rangle
-\langle \Upsilon^{(w+{c\over p})}_{J_c}|\partial_l
\Upsilon^{(w+{c\over p})}_{J_c}\rangle
\big)
\nonumber\\
&&
+\big(
\langle\partial_l \Upsilon^{(w+{c\over p})}_J|
\Upsilon^{(w+{c\over p})}_J\rangle
-\langle \Upsilon^{(w+{c\over p})}_J|\partial_l
\Upsilon^{(w+{c\over p})}_J\rangle
\big)
\big]
\Big\}.
\label{paper4.57}
\end{eqnarray}
After the sum over the sub-bands of a cluster one gets
a sum rule
\begin{eqnarray}
\sigma_{xy}=
\sum_{\cal S}'{\cal K}_{{\cal S}\sigma} \qquad {\cal K}_{{\cal S}\sigma}
\equiv \sum_{J\in{\cal S }}k_{J\sigma}.
\label{paper4.58}
\end{eqnarray}
\par
The last results, together with eq. (\ref{paper4.41e}), implies
\begin{eqnarray}
{1\over p}[{\cal N}_{\cal S} - {{q{\cal K}'_{\cal S}}\over{p}}] =  {\cal K}_{{\cal S}\sigma} 
\label{paper4.58a}
\end{eqnarray}
for every cluster.
Since $p$ and $q$ are relative prime numbers, then 
\begin{eqnarray}
{\cal M}_{\cal S}\equiv {{{\cal K}'_{\cal S}}\over{p}}  
\label{paper4.58b}
\end{eqnarray}
must be an integer and thus we get the Diofantine equation
\begin{eqnarray}
p~ {\cal K}_{{\cal S}\sigma} + q~ {\cal M}_{\cal S} = {\cal N}_{\cal S}
\label{paper4.59}
\end{eqnarray}
where ${\cal N}_{\cal S}$ is the number of sub-bands in the cluster.
The validity of the result is general: only the gap condition is required.
Once this condition  is satisfied, eq. 
(\ref{paper4.59}) should be valid independently from the strength
of the periodic potential.
\section{Adiabatic approximation}
In the adiabatic approximation ($\tau \to \infty$) the Floquet
state is the instantaneous eigenvector of the Hamiltonian
\begin{equation}
H^{(\Upsilon)}(w+{\tilde{\cal E}}t)|\Upsilon_J^{(w)}(t)\rangle
=\lambda_J(w+{\tilde{\cal E}}t)|\Upsilon_J^{(w)}(t)\rangle
\label{paper4.42}
\end{equation}
were the phase is fixed by the Schr\"odinger equation
projected on the vector $\Upsilon_J^{(w)}(t)$
\begin{eqnarray}
\langle\Upsilon_J^{(w)}(t)|\partial_t\Upsilon_J^{(w)}(t)\rangle
&=& -i\langle\Upsilon_J^{(w)}(t)|H^{\Upsilon}(w+{\tilde{\cal E}}t)
|\Upsilon_J^{(w)}(t)\rangle
\nonumber\\
&=& -i \lambda_J(w+{\tilde{\cal E}}t).
\label{paper4.43}
\end{eqnarray}
The dependence of $\Upsilon$ from $w$ and $t$ is through
the combination $w+{\tilde{\cal E}}t$.
\par
The above equation is very important for numerical computation.
The phase of the eigenvectors yielded by a computer is usually
fixed by requiring that the largest component is real. This 
choice does not satisfy in general the necessary continuity
requirement. Eq. (\ref{paper4.43}) allows to fix the phase
in the correct way. First one makes a choice of a regular phase on two
side vectors of the reduced Brillouin zone. Then the phase of
the wave function $\Upsilon_J^{(w)}(t)$ is fixed over the whole
reduced Brillouin zone by imposing condition (\ref{paper4.43})
over the straight lines parameterized by $w+{\tilde{\cal E}}t,~
t\in {\cal R}$.
\par
The Floquet energies are given by
\begin{equation}
E_J^{(w)} = \int_0^\tau  dt \lambda_J(w+{\tilde{\cal E}}t)
+ \alpha_J^{(w)}
\label{paper4.44p}
\end{equation}
where $\alpha_J^{(w)}$ is the Berry phase. 
For isolated eigenvalues the adiabatic energy is
expected to be periodic over the reduced Brillouin zone
\begin{equation}
\lambda_J(w+ m{c\over p}+n{d\over p})
=\lambda_J(w)
\label{paper4.45}
\end{equation}
then eq. (\ref{paper4.37}) requires
\begin{eqnarray}
&&
\alpha^{(w+{c\over p})}_J =
\alpha^{(w)}_J + 2\pi k_{Jc}
\nonumber\\
&&
\alpha^{(w+{d\over p})}_J =
\alpha^{(w)}_J + 2\pi k_{Jd}.
\label{paper4.46}
\end{eqnarray}
By a similar argument one gets also
\begin{equation}
{\partial\over {\partial t}}
\alpha^{(w+{\tilde{\cal E}} t)}_J = 0.
\label{paper4.47}
\end{equation}
%
\section{Example}
We consider the potential
\begin{eqnarray}
V({\bf r})=v_1\cos({{q}\over{p}}{\tilde{\bf c}'}
\cdot{\bf r}) + v_2\cos({{q}\over{p}}{\tilde{\bf d}'}
\cdot{\bf r}).
\label{paper4.51}
\end{eqnarray}
This potential is particularly simple since the
exponential can be written as
\begin{eqnarray}
\exp[i {\tilde {\bf v}}\cdot{\bf r}]
= S(v)T(-v)
\label{paper4.52}
\end{eqnarray}
and therefore the cosine function has a simple expression in terms
of unitary operators $S$ and $T$ (eqs. (\ref{paper4.6})
and (\ref{paper4.11})). The basis is given by
\begin{eqnarray}
\Phi^{\alpha\beta}_{n_L} = S(w_{\alpha\beta})
\Phi^{\alpha_0\beta_0}_{n_L} \qquad n_L = 0,\dots,n_M
\label{paper4.53}
\end{eqnarray}
where $n_L$ is Landau level number and
$\alpha,\beta$ have $p$ values and 
are the pseudo-momenta associated
to a finer tiling ($\bf f,g$) (with flux one) of ($\bf c,d$) lattice.
They are fixed by the boundary conditions (\ref{paper4.10}).
Moreover 
\begin{equation}
w_{\alpha\beta} = {1\over{2\pi}}
[(\beta-\beta_0)f   -(\alpha-\alpha_0) g].
\label{paper4.54}
\end{equation}
The size of the basis is fixed by $n_M$, that has to be chosen
large enough ($(n_M+{1\over 2})>> |{\cal V}|$).
\par
With the given basis one evaluates $U(\tau,0)$.  One
extracts the Floquet energies and states by diagonalization 
of $U(\tau,0)$.
\par
The direction of the field is chosen to be
\begin{equation}
\tau{\tilde{\cal E}}= 2 {\bf c}' + {\bf d}'  
\label{paper4.55a}
\end{equation}
as a compromise between generality and simplicity.
The choice of
the direction of the electric field determines the periodicity
patterns for the mean energy and for the eigenvalues of the evolution
operator $U(\tau,0)$ (or equivalently for the Floquet energies
mod$(2\pi)$).
\par
The figures 1-8 show the changes in the mean energy (\ref{paper4.41ab})
and in the Floquet energy by varying the period $\tau$
and consequently the electric field, according to eq. (\ref{paper4.55a}).
In the whole set of examples we consider the case $q=2, p=3$ with
\begin{equation}
\left \{
\begin{array}{l}
{\bf c}={\bf c}'-{\bf d}'
\\
{\bf d}={\bf c}' + {\bf d}'
\end{array}
\right .
\label{paper4.55b}
\end{equation}
and use $v_1=v_2=0.5$. We take a basis
with Landau number $n_L=0,\dots,4$. We show here only the three
lowest sub-bands. The relevant factor in eq. (\ref{paper4.41}) is
then
\begin{eqnarray}
\tau{\bf c}\cdot {\cal E} = 2\pi{{9}\over{2}}.
\label{paper4.55c}
\end{eqnarray}
\par
The abscissa is a coordinate of the reduced Brillouin zone along
the electric field (in the orthogonal direction everything is
constant, see eq. (\ref{paper4.40})). The $x=1.0$ point corresponds
to $w={{c}\over{p}}$.
\par
Fig. 1 and 2 give the mean energy and the Berry phase in the adiabatic
approximation (i.e. we drop in eq. (\ref{paper4.44p})
the irrelevant part coming from the instantaneous
eigenvalue). The sub-band denoted by circles has 
$(2\pi)^{-1}\Delta E=-3$ and therefore $\sigma_{\rm H}=1$. Stars give
$(2\pi)^{-1}\Delta E=6$ and  $\sigma_{\rm H}=-1$. Boxes give
$(2\pi)^{-1}\Delta E=-3$ and  $\sigma_{\rm H}=1$.
\par
As the period  decreases ($\tau=400$ for Figs. 3 and 4)
the gap between the two lowest sub-bands disappears and the periodicity
(of mean energy and Floquet energy mod$(2\pi)$) is lost. The cluster
given circles and stars give together $(2\pi)^{-1}\Delta E=3$ and 
$\sigma_{\rm H}=0$. Boxes give
$(2\pi)^{-1}\Delta E=-3$ and  $\sigma_{\rm H}=1$.
\par
Figs. 5 and 6 catch the moment ($\tau=300$) where also the third
sub-band start crossing the other two. Fig. 9 shows what happens
to the Floquet energies: as the period increases 
the lines approach each other wedge-wise and eventually merge to form
two (almost straight) crossing lines. 
\par
Figs. 7 and 8 show the situation at the shorter period $\tau=100$.
By decreasing further the period higher lying sub-bands cross
the considered set of three.
\par
For both $\tau=300$ and $\tau=100$ the cluster is formed by the
three sub-bands: $(2\pi)^{-1}\Delta E=0$ and  $\sigma_{\rm H}=1$.
\par
We conclude the section with few comments. 
\begin{itemize}
\item The relation between period and strength of electric field
depends on the lattice site as in eq. (\ref{paper4.22}). Then one
should ask which parameter is relevant for the structure of the
sub-bands. We have done few numerical experiments, by keeping
the period fixed and 
by varying the lattice site (thus changing the 
strength of the electric field). It turns out
that the relevant parameter is the strength of the electric field.
The change of the lattice site amounts to a change in the periodicity
pattern, but the cluster structure remains the same. Fig. 10
provides an instance of this search for the case $\tau =300$,
$\tau{\tilde{\cal E}}= 5 c' + d'$.
\item Consider the situation described by the Figs. 5 and 6. What
happens when the chemical potential is $\simeq 0.6$ i.e. in the
almost open gap? The condition in eq. (\ref{paper4.23}) excludes the
parts of the sub-bands above the chemical potential. Thus the filling
factor is $\nu_f \simeq 2/3$. An inspection of Fig. 6 shows
that $(2\pi)^{-1}\Delta E\simeq 3$ and therefore 
$\sigma_{\rm H}\simeq0$ as in the case described by Figs. 3 and 4.
However, since there is no gap, the longitudinal conductivity
should be appreciably different from zero.
\item We have considered a larger basis ($n_M=12$) in the
adiabatic approximation. 
It is intriguing that for each set of sub-bands coming
from the same Landau level the contribution to the transverse Hall
conductivity has the  pattern (1,-1,1), but it has an exceptional
behavior for $n_L=3$ where it is (-1,3,-1). We have no clues for this
anomaly.
\end{itemize}
\section{Acknowledgment}
We acknowledge a partial financial support by MURST. We wish to thank
Luca Molinari for useful discussions.

%
\begin{figure}
\centerline{
\epsfysize = 8cm
\epsffile{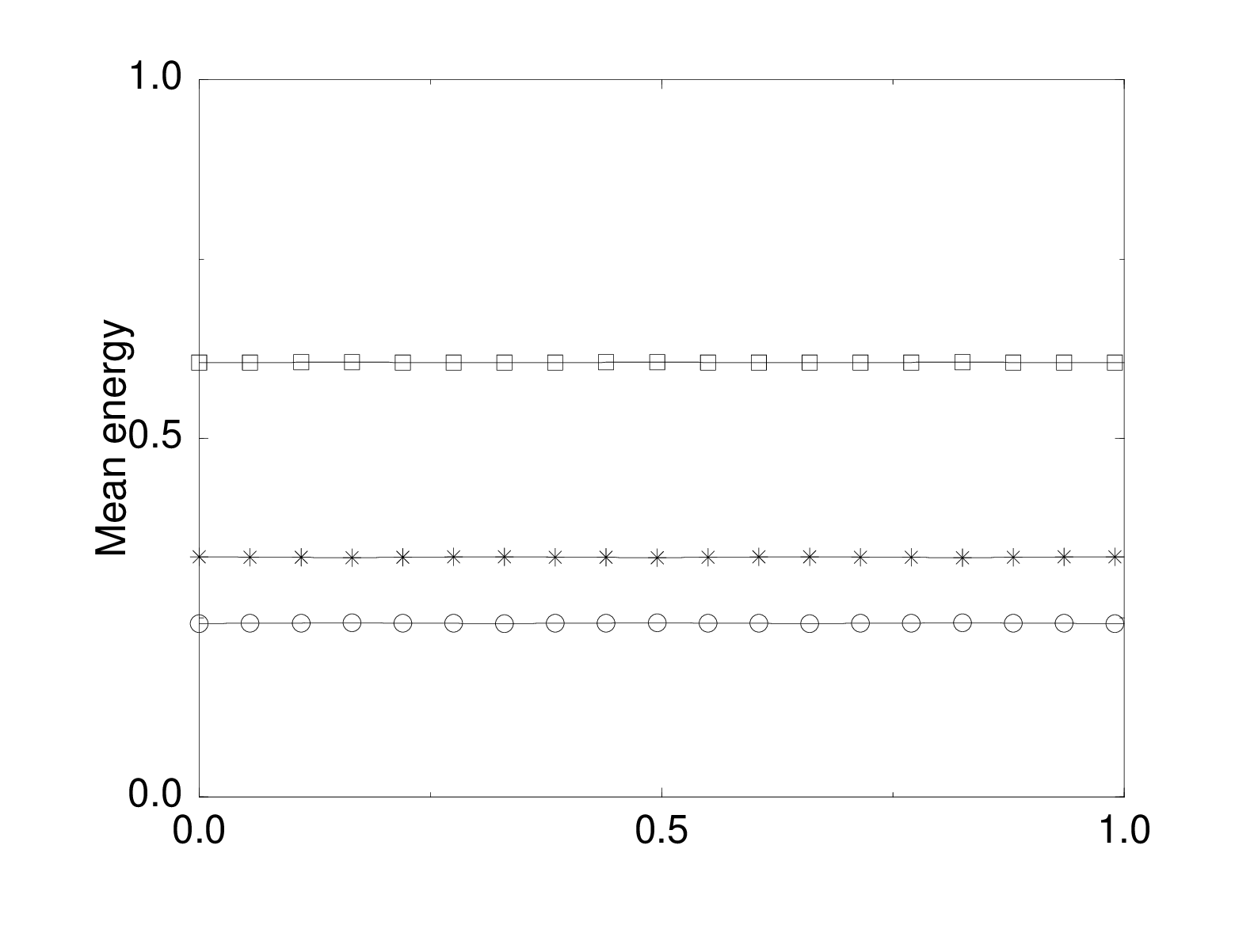}}
\caption{Mean energy in the adiabatic approximation \label{adiae}}
%
\centerline{
\epsfysize = 8cm
\epsffile{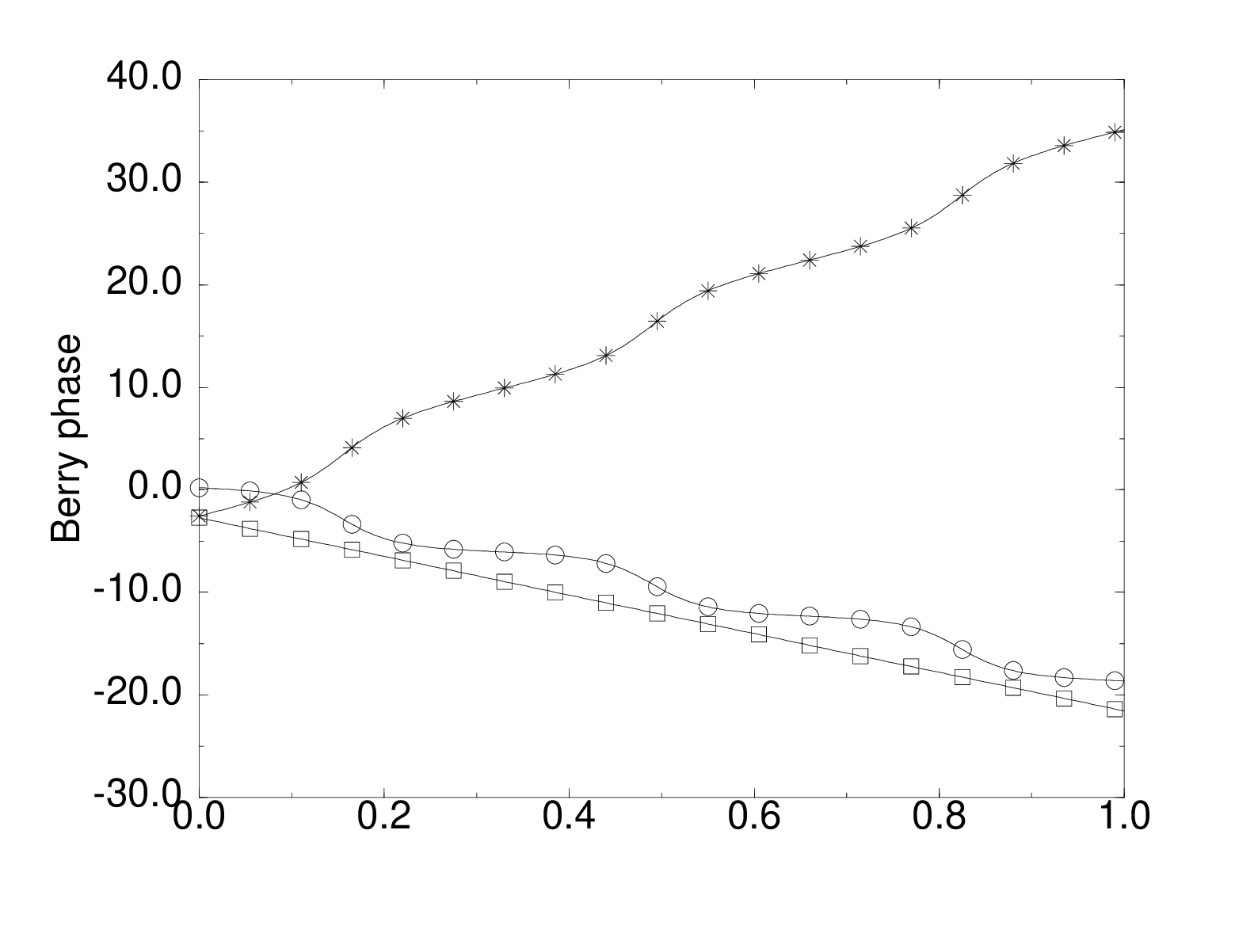}}
\caption{ Berry phase in the adiabatic approximation. \label{adiaf}}
\end{figure}
\begin{figure}
\centerline{
\epsfysize = 8cm
\epsffile{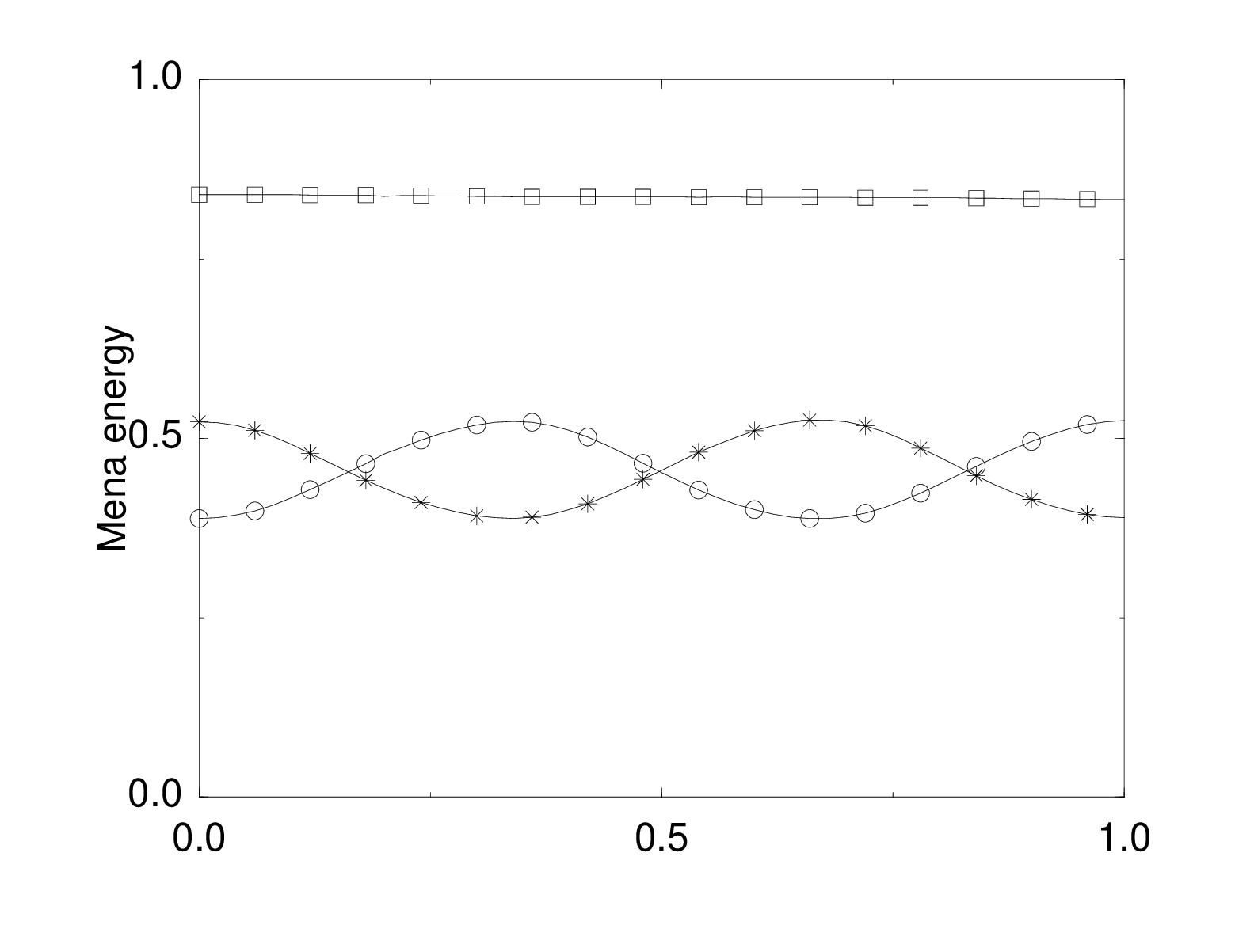}}
\caption{ Mean energy for period $\tau=400$. \label{400e}}
%
\centerline{
\epsfysize = 8cm
\epsffile{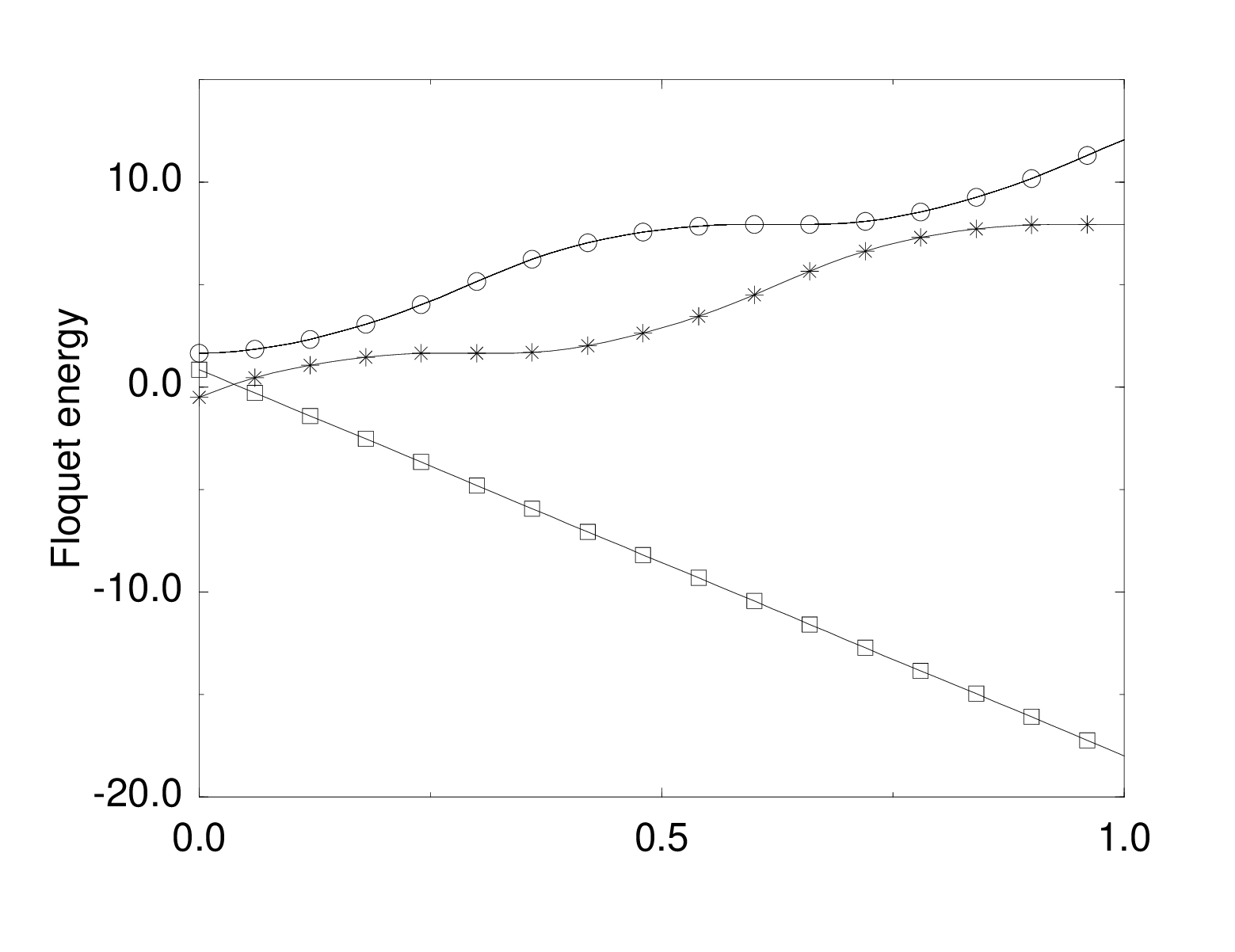}}
\caption{ Floquet energy for period $\tau=400$. \label{400f}}
\end{figure}
\begin{figure}
\centerline{
\epsfysize = 8cm
\epsffile{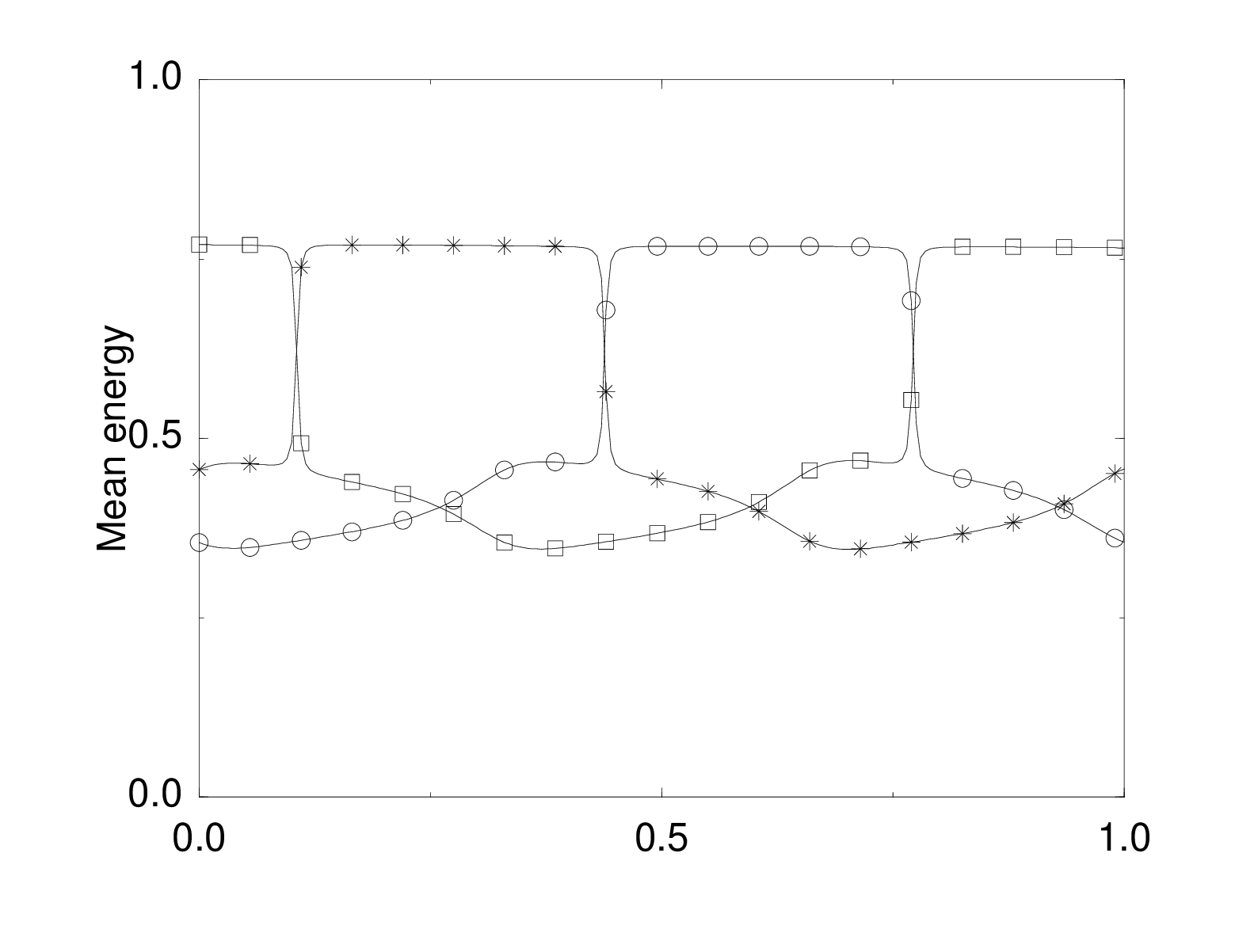}}
\caption{Mean energy for period $\tau=300$. \label{200e}}
%
\centerline{
\epsfysize = 8cm
\epsffile{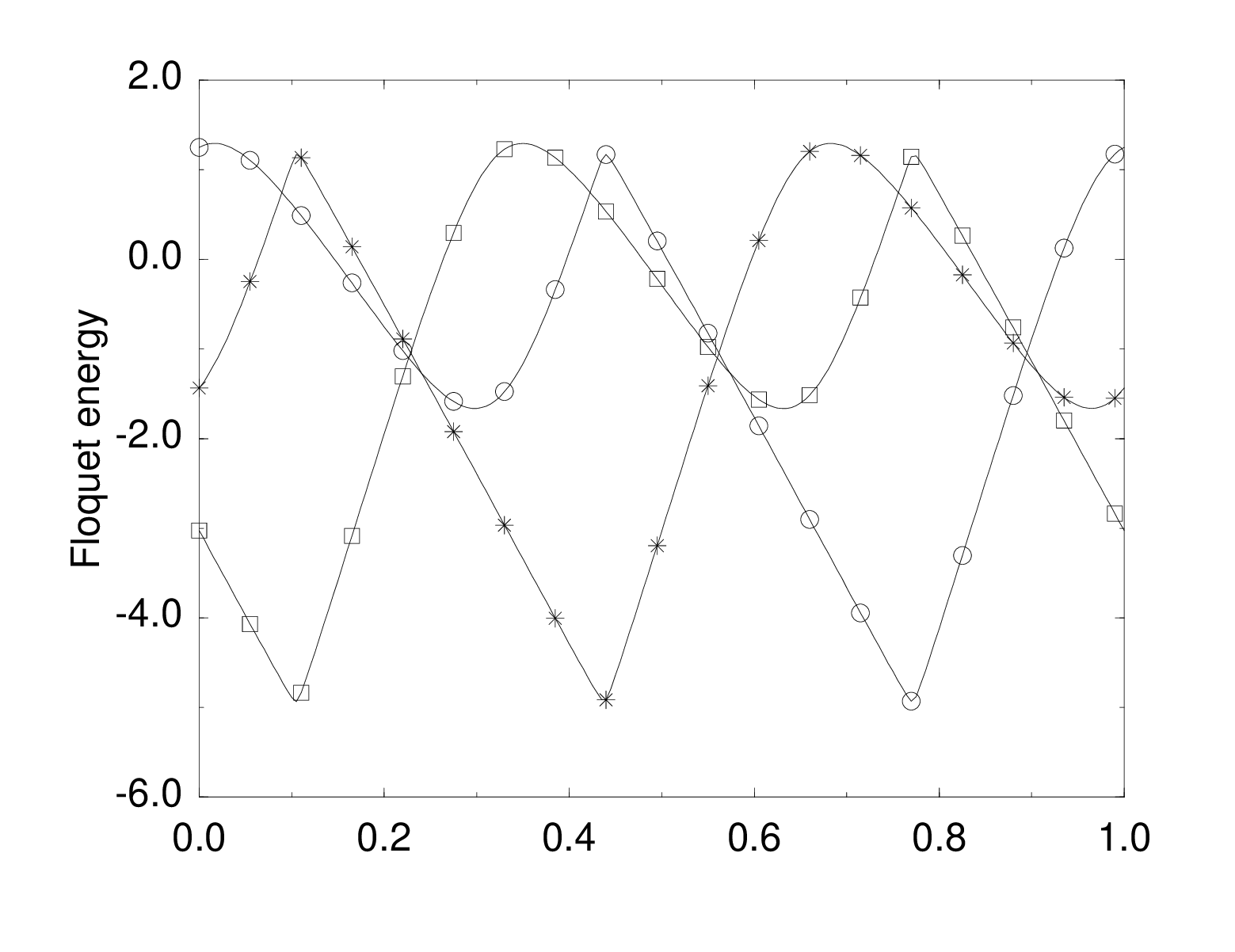}}
\caption{Floquet energy for period $\tau=300$. \label{200f}}
\end{figure}
\begin{figure}
\centerline{
\epsfysize = 8cm
\epsffile{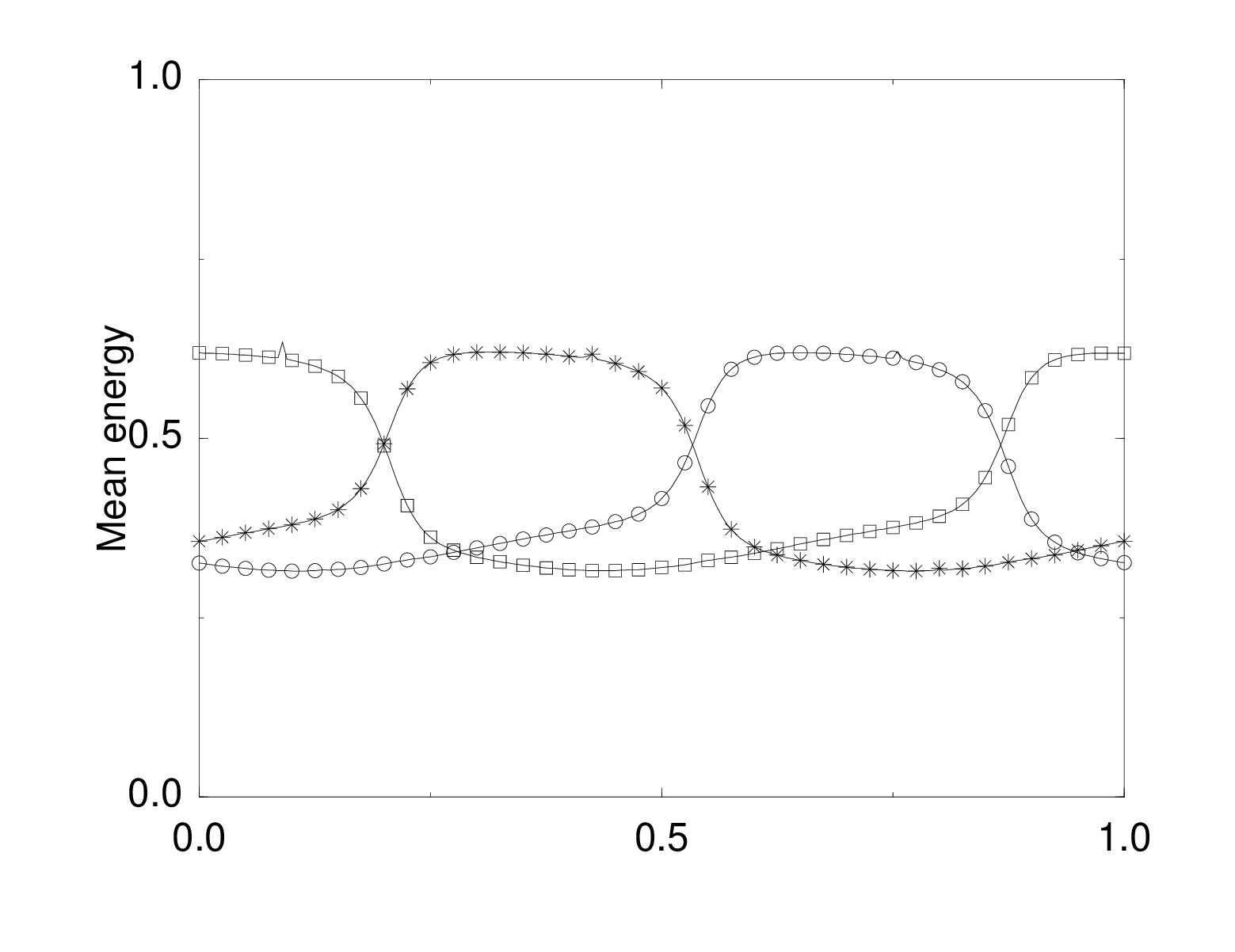}}
\caption{Mean energy for period $\tau=100$. \label{100e}}
%
\centerline{
\epsfysize = 8cm
\epsffile{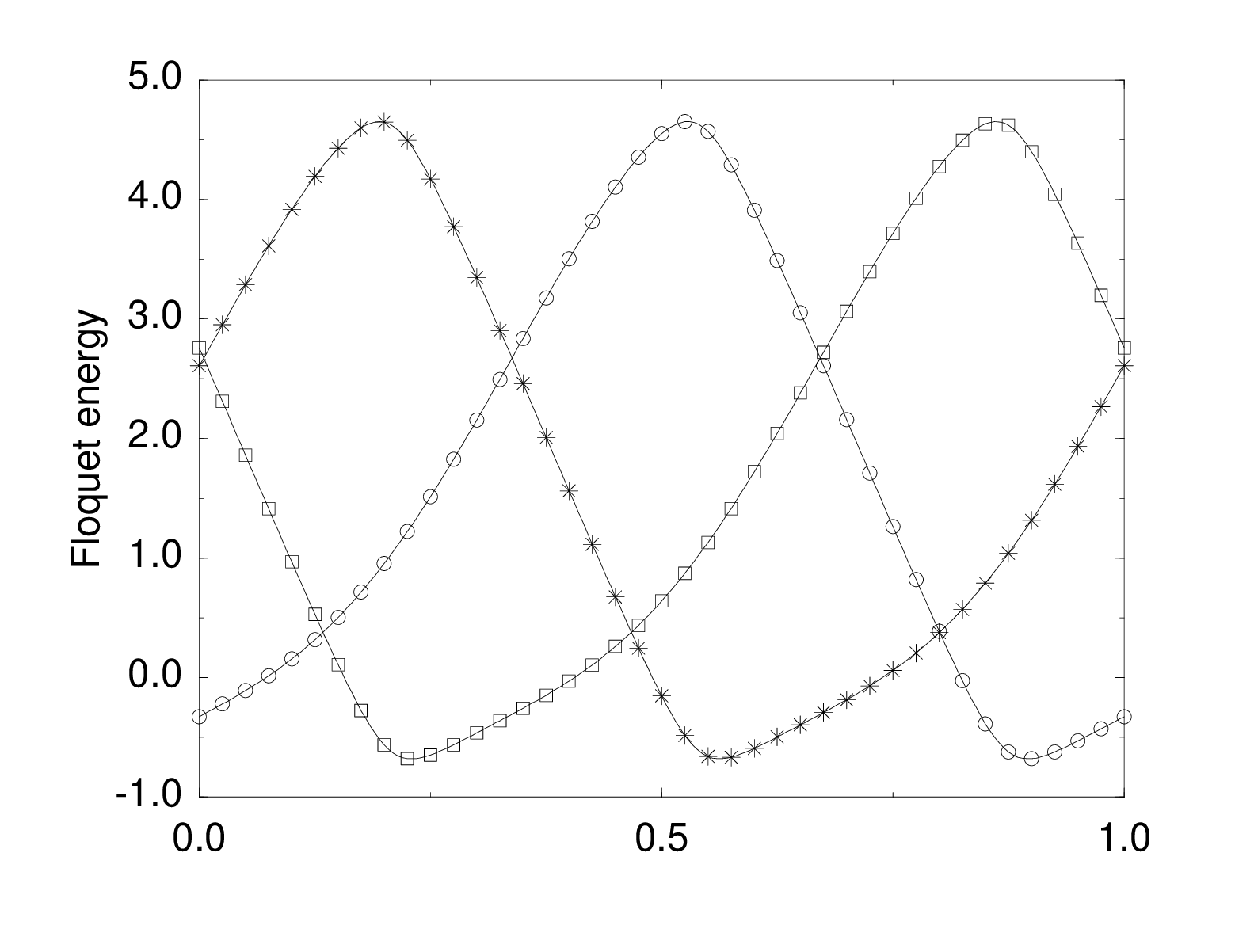}}
\caption{ Floquet energy for period $\tau=100$. \label{100f}}
\end{figure}
\begin{figure}
\centerline{
\epsfysize = 8cm
\epsffile{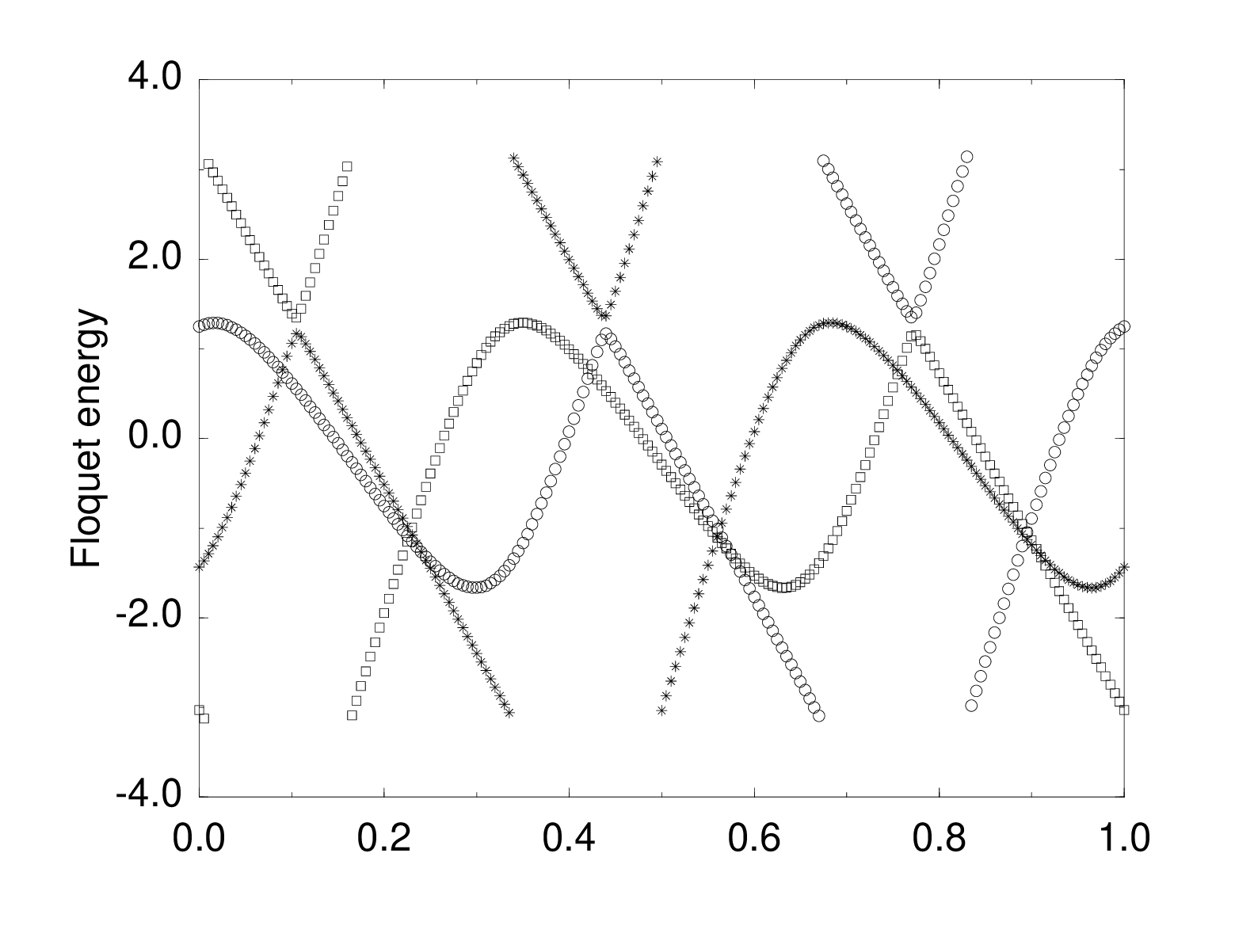}}
\caption{Floquet energy (mod$2\pi$) for period $\tau=300$. \label{300fp}}
\end{figure}
\begin{figure}
\centerline{
\epsfysize = 8cm
\epsffile{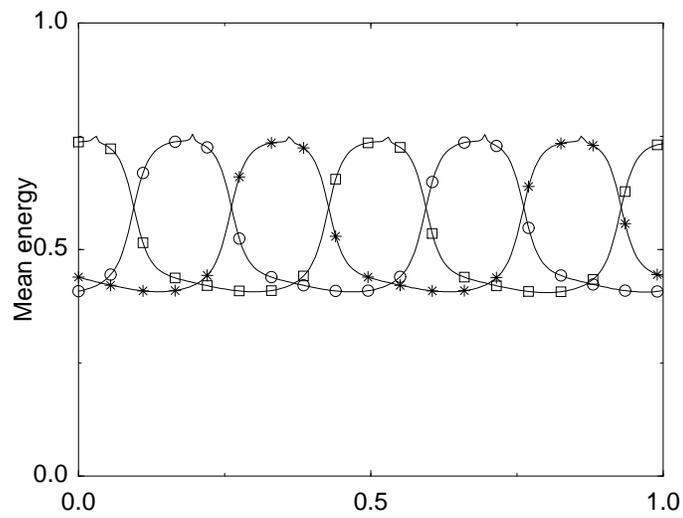}}
\caption{Mean energy for period $\tau=300$, 
$\tau{\tilde{\cal E}}=5~c'+ d'$. \label{300fpx}}
\end{figure}
\vfil\eject
\end{document}